\documentclass[conference]{IEEEtran}
\IEEEoverridecommandlockouts
\usepackage[utf8]{inputenc}
\usepackage{cite}
\usepackage{overpic}
\usepackage{amsmath,amssymb,amsfonts}
\usepackage{bm}
\usepackage{algorithmic}
\usepackage{graphicx}
\usepackage{textcomp}
\usepackage{xcolor}
\usepackage{float}
\usepackage{amsthm}
\usepackage{graphicx}
\usepackage{epstopdf}
\usepackage{amsmath,bm,bbm}
\usepackage{amsfonts}
\usepackage{amssymb}
\usepackage{color}
\usepackage{multirow}
\usepackage{multicol}
\usepackage{soul,xcolor}
\usepackage{setspace}

\usepackage{longtable}
\usepackage{mathtools}
\usepackage{subfig}
\usepackage{tabulary}
\usepackage{epsfig}
\usepackage{caption}
\usepackage{booktabs}
\usepackage{blindtext}
\usepackage{adjustbox}
\usepackage{hyperref}
\usepackage{dirtytalk}
\usepackage{comment} 

\usepackage{tikz}
\usetikzlibrary{positioning}

\definecolor{mypurple}{HTML}{9933FF}
\definecolor{mygreen}{HTML}{009900}

\DeclareMathOperator{\tr}{tr}

\theoremstyle{plain}

\newcommand{\mathacr}[1]{\mathsf{#1}}

\newcommand{\vect}[1]{\mathbf{#1}}

\def\tr{\mathrm{tr}}

\def\Htran{\mbox{\tiny $\mathrm{H}$}}
\def\Ttran{\mbox{\tiny $\mathrm{T}$}}

\begin{document}
\bstctlcite{IEEEexample:BSTcontrol}
\makeatletter
\newcommand*{\rom}[1]{\expandafter\@slowromancap\romannumeral #1@}
\makeatother

\title{Energy-Aware Resource Allocation for Multi-Operator Cell-Free Massive MIMO in V-CRAN Architectures}
\vspace{-1cm}

\author{

 \IEEEauthorblockN{Derya Nurcan-Atceken$^*$, \"Ozlem Tu\u{g}fe Demir$^\dagger$, Aysegul Altin-Kayhan$^*$, Emil Bj{\"o}rnson$^\ddagger$, Cicek Cavdar$^\ddagger$, Bulent Tavli$^\dagger$}
\IEEEauthorblockA{ {$^*$Department of Industrial Engineering, TOBB ETÜ, Ankara, Turkiye}
\\ 
 {$^\dagger$Department of Electrical-Electronics Engineering, TOBB ETÜ, Ankara, Turkiye}
\\
 {$^\ddagger$Department of Computer Science, KTH Royal Institute of Technology, Stockholm, Sweden} \\
		{Email: deryanurcan@etu.edu.tr, ozlemtugfedemir@etu.edu.tr, aaltin@etu.edu.tr, emilbjo@kth.se, cavdar@kth.se, btavli@etu.edu.tr }
}
\vspace{-1cm}

\thanks{Derya Nurcan-Atceken was supported by TÜBİTAK BİDEB 2211 Scholarship Programme, Turkiye. The work by \"O. T. Demir was supported by 2232-B International Fellowship for Early Stage Researchers Programme funded by the Scientific and Technological Research Council of Turkiye. E.~Bj\"ornson was supported by the SUCCESS grant from SSF.}

}

\maketitle
\begin{abstract}
Cell-free massive multiple-input multiple-output (MIMO) implemented in virtualized cloud radio access networks (V-CRAN) has emerged as a promising architecture to enhance spectral efficiency (SE), network flexibility, and energy efficiency (EE) in next-generation wireless systems. In this work, we develop a holistic optimization framework for the efficient deployment of cell-free massive MIMO in V-CRAN with multiple mobile network operators (MNOs). Specifically, we formulate a set of integer programming models to jointly optimize access point (AP) selection, user equipment (UE) association, cloud resource allocation, and MNO assignment while minimizing the maximum total power consumption (TPC) across MNOs. We consider two different scenarios based on whether UEs can be assigned to arbitrary MNOs or not. The numerical results demonstrate the impact of different deployment assumptions on power consumption, highlighting that flexible UE-MNO assignment significantly reduces TPC. The findings provide key insights into optimizing resource management in cell-free massive MIMO V-CRAN, paving the way for energy-efficient wireless network implementations.
\end{abstract}

\vspace{-2mm}

\section{Introduction}
\vspace{-1mm}

The rapid growth of mobile data traffic and the increasing demand for seamless connectivity are accelerating the evolution of next-generation wireless networks \cite{han2018network, ngo2024ultradense}. Massive MIMO (multiple-input multiple-output) has become a cornerstone of this evolution, offering substantial gains in spectral efficiency (SE) compared to conventional small-scale MIMO systems by enabling spatial multiplexing and precise beamforming with large antenna arrays \cite{massivemimobook}. Yet, traditional cellular architectures face significant performance degradation at cell edges due to inter-cell interference.

To address these limitations, cell-free massive MIMO has been proposed as a post-cellular solution that replaces the large cellular base stations with a large number of distributed, low-cost access points (APs) cooperatively serving user equipments (UEs) on the same time-frequency resources \cite{Ngo2017b, interdonato2019ubiquitous, cell-free-book}. Despite each AP having only a few antennas, the system as a whole maintains the “massive” property due to the high aggregate antenna count.

 The APs are coordinated by central processing units (CPUs), linking cell-free architectures to emerging cloud-based radio access networks such as C-RAN and virtualized C-RAN (V-CRAN) \cite{wang2020implementation,demir2024cell}. Unlike traditional C-RAN, where each radio unit (RU) retains a fixed processing unit and fronthaul fiber resources, V-CRAN decouples hardware and software via network function virtualization, enabling dynamic resource sharing across multiple nodes \cite{wang2017virtualized, wang2016joint, wang2016energy, Masoudi2020, sigwele2017energy, masoudi2019green}. This enhances overall energy efficiency (EE) by enabling virtualization-driven consolidation of processing tasks on general-purpose processors (GPPs), allowing for flexible workload distribution, hardware reuse, and reduced idle power consumption across the network. The recent work \cite{demir2024cell} highlights the importance of jointly managing radio, fronthaul, and processing units to unlock the full energy-saving potential of cell-free V-CRAN systems.

\vspace{-1mm}

\subsection{Related Work}

\vspace{-1mm}

Most prior works on cell-free massive MIMO have primarily focused on optimizing radio transmit power, while keeping all hardware turned on. However, as highlighted in \cite{demir2022cell,demir2024cell}, it is critically important to consider the end-to-end power consumption and the allocation of virtualized resources to fully exploit the energy-saving potential of the V-CRAN architecture. Notably, radio transmit power—when considered in isolation from idle, fronthaul, and processing power—accounts for only a small fraction of the total end-to-end energy consumption. 

In this paper, we build upon \cite{demir2022cell,demir2024cell} by incorporating multiple physical mobile network operators (MNOs), each operating over distinct frequency bands and managing its own radio, fronthaul, and cloud infrastructure without inter-operator cooperation. The UE–MNO assignment is either fixed or optimized depending on the scenario, but each UE is exclusively served by one MNO's infrastructure. Our goal is to answer the following research question: Can flexible UE–MNO assignment, where each UE is dynamically associated with a single MNO's infrastructure, provide any energy savings?

Cooperation among MNOs in terms of passive and active infrastructure sharing, spectrum usage, and operational responsibilities has been discussed in \cite{markendahl2013network}. Furthermore, \cite{fander2022game} analyzed MNO cooperation from the perspective of infrastructure cost-sharing under government intervention. However, neither of these works considered cell-free massive MIMO nor the concept of flexible UE-MNO assignment.

\vspace{-1mm}

\subsection{Contributions}
\vspace{-1mm}

This paper advances the state-of-the-art in cell-free massive MIMO over V-CRAN through the following contributions:
\begin{itemize}
\item We propose two deployment scenarios with varying degrees of flexibility in UE–MNO associations. For each scenario, we formulate and solve a novel 0–1 integer programming model that jointly optimizes UE-to-AP assignment, active AP selection, and MNO-specific provisioning of cloud and fronthaul resources while minimizing the maximum total power consumption (TPC) across MNOs.
\item We show that the SE constraints in the power minimization problem involve quadratic terms, which could in principle be handled using second-order cone (SOC) formulations. However, due to the 0–1 binary nature of the AP–UE assignment and activation variables, we propose a tailored linearization technique that enables efficient solution using 0–1 linear programming solvers, significantly reducing computational complexity.

\end{itemize}

\section{System Model}

We consider a cell-free massive MIMO system with $M$ MNOs, each deploying a V-CRAN architecture as illustrated in \cite[Fig.~1]{demir2022cell}. Each MNO operates in a distinct frequency band, eliminating inter-operator interference. The network comprises a total of $L$ APs and $K$ single-antenna UEs, all arbitrarily distributed over the coverage area. Each AP is equipped with $N$ antennas and connects to its respective MNO’s cloud via fronthaul links, forming a V-CRAN as described in \cite{wang2017virtualized}.

We assume that the AP-to-MNO assignment is fixed; however, we consider two scenarios: one where UEs are flexibly assigned to any MNO (along with its associated radio, fronthaul, and cloud resources), and another where each UE is statically bound to a single MNO. A subset of APs is selected to serve each UE in order to satisfy its SE requirement. We let $x_{kl} \in \{0,1\}$ denote whether AP $l$ serves UE $k$. These associations are optimized to minimize the maximum TPC across all MNOs while ensuring SE constraints are met.

We adopt the eCPRI standard with split E \cite{perez2018fronthaul}, where APs perform only RF tasks, while all baseband processing is centralized in the digital unit (DU) cloud. Fronthaul transport uses time-wavelength division multiplexing passive optical network (TWDM-PON) to carry eCPRI packets \cite{wang2016energy,wang2017virtualized}. Each AP connects to an optical network unit (ONU) on a shared wavelength. In the cloud, an optical line terminal (OLT) with WDM MUX and $W$ stacks of DUs routes data accordingly, with each DU connected to one wavelength via a line card (LC). GPPs are assumed for virtualized baseband processing. 

In the following, we focus on a particular MNO indexed by $m$. Let $\mathcal{A}_m$ and $\mathcal{U}_m$ denote the sets of APs and UEs associated with MNO $m$, for $m = 1, \ldots, M$.

We assume time-division duplex operation with frequency-selective channels. Under the block fading assumption, it is assumed that the channels are constant across $\tau_c$ time-frequency channel uses in each coherence block. For the other details, please refer to \cite{demir2024cell}. The system operates in distributed downlink mode, using local channel estimates for per-AP precoding \cite[Sec. 6.2]{cell-free-book}, enabling virtualization and reduced DU usage.

Each coherence block is split into $\tau_p$ uplink training and $\tau_d = \tau_c - \tau_p$ downlink data samples. All UEs are served simultaneously via spatial multiplexing, and we focus on a representative time-frequency block as in \cite{cell-free-book}.

\subsection{Downlink Spectral Efficiency}

We let $\vect{h}_{kl} \in \mathbb{C}^{N}$ denote the channel from UE $k$ to AP $l$ within an arbitrary coherence block. The channels are modeled as correlated Rayleigh fading, following the framework in \cite{demir2024cell}. We assume channel estimation is performed using the minimum mean-squared error (MMSE) method. Each AP employs local partial-MMSE (LP-MMSE) precoding, and $\vect{w}_{kl}\in \mathbb{C}^{N}$ denotes the precoding vector applied by AP $l$ for UE $k$, as described in \cite{demir2024cell}. The achievable downlink SE for UE $k$—a lower bound on the ergodic capacity obtained via the use-and-then-forget (UatF) bound—can be computed using  \cite[Corr.~6.3 and Sec. 7.1.2]{cell-free-book} as
	\begin{equation} \label{eq:downlink-rate-expression-level2}
	\mathacr{SE}_{k} = \frac{\tau_d}{\tau_c} \log_2  \left( 1 + \mathacr{SINR}_{k}  \right) \quad \textrm{bit/s/Hz}
	\end{equation}
	with the effective signal-to-interference-plus-noise ratio (SINR) given by
\begin{equation}\label{eq:SINR-downlink-2}
\mathacr{SINR}_{k}=\frac{\left|\vect{b}_k^{\Ttran}{\boldsymbol{\rho}}_k\right|^2}{\sum\limits_{i\in \mathcal{U}_m}{\boldsymbol{\rho}}_i^{\Ttran}{\vect{C}}_{ki}{\boldsymbol{\rho}}_i-\left|\vect{b}_k^{\Ttran}{\boldsymbol{\rho}}_k\right|^2+\sigma^2},
\end{equation}
where
\begin{align}
 &{\boldsymbol{\rho}}_k=\left [ \sqrt{p_{k1}}x_{k1} \, \ldots \, \sqrt{p_{kL}} x_{kL} \right ]^{\Ttran} \in \mathbb{R}_{\geq 0}^L,  \\
  &{\vect{b}}_k \in \mathbb{R}_{\geq 0}^L, \quad
    \left[ {\vect{b}}_{k} \right]_{l}= \mathbb{E} \left\{ \vect{h}_{kl}^{\Ttran}{\vect{w}}_{kl}\right\},\\
  & {\vect{C}}_{ki}\in \mathbb{C}^{L \times L}, \nonumber \\
  & \left[ {\vect{C}}_{ki} \right]_{lr}=
  \mathbb{E} \left\{ \vect{h}_{kl}^{\Ttran}{\vect{w}}_{il}{\vect{w}}_{ir}^{\Htran}\vect{h}_{kr}^*\right\}.
\end{align}
For notational simplicity, the summation includes all AP indices; however, for APs that do not belong to the same operator as the UE, the corresponding power coefficients $p_{kl}$ are set to zero. The noise variance is denoted by $\sigma^2$ in \eqref{eq:SINR-downlink-2}.

In the considered quality-of-service (QoS)-aware optimization problems, each UE will have a specific SE threshold, which is equivalent to the constraint $\mathrm{SINR}_k\geq \gamma_k$ for some minimum SINR threshold $\gamma_k$.  To make the considered optimization problems tractable, we consider a fixed power allocation strategy.  The SINRs can then be written in terms of the vectors
\begin{align}
\vect{x}_k = [x_{k1} \ \ldots \ x_{kL}]^{\Ttran}, k\in \mathcal{U}_m.
\end{align}
We define
\begin{align}
\vect{p}_k=\left [ \sqrt{p_{k1}} \, \ldots \, \sqrt{p_{kL}} \right ]^{\Ttran} \in \mathbb{R}_{\geq 0}^L
\end{align}
and 
\begin{align}
    \overline{\vect{C}}_{ki}= \begin{cases} \frac{\vect{C}_{kk}\odot(\vect{p}_k\vect{p}_k^{\Ttran})}{\sigma^2}-\frac{\gamma_k+1}{\gamma_k\sigma^2}(\vect{b}_k\vect{b}_k^{\Htran})\odot(\vect{p}_k\vect{p}_k^{\Ttran}),  & i=k , \\\frac{\vect{C}_{ki}\odot(\vect{p}_i\vect{p}_i^{\Ttran})}{\sigma^2}, & i\neq k.
    \end{cases}
\end{align}
Then, the SINR constraint $\mathrm{SINR}_k\geq \gamma_k$ can be written as
\begin{align}
 \sum_{i\in \mathcal{U}_m}\tr\left(\overline{\vect{C}}_{ki}\vect{x}_i\vect{x}_i^{\Ttran}\right)     +1 \leq 0.
\end{align}
We define $a_{ilr}=x_{il}x_{ir}$ and apply linearization to the multiplication of binary variables using the constraints
\begin{align}
&a_{ilr}\leq x_{il},\\
&a_{ilr}\leq x_{ir}, \\
& a_{ilr}\geq x_{il}+x_{ir}-1, \\
& a_{ilr}\in \{0,1\}.
\end{align}
Denoting the $(l,r)$th entry of $\overline{\vect{C}}_{ki}$ as $\overline{C}^k_{ilr}$, we obtain the SINR constraint in linear form as
\begin{align}
 \sum_{i\in \mathcal{U}_m}\sum_{l\in \mathcal{A}_m}\sum_{r\in \mathcal{A}_m}\overline{C}^k_{ilr}a_{ilr}    +1 \leq 0.
\end{align}

\section{Service Provisioning Scenarios} \label{subsec:scenarios}

In this subsection, we provide a detailed explanation of the scenarios we consider under the assumption that the UEs are either known or unknown in terms of which service provider cloud they are connected to. Each MNO maintains its own cloud and fronthaul network, operating independently without any cooperation between them. Based on these assumptions, we define the following two conditions that apply across all scenarios:
\vspace{-0.5mm}
\begin{enumerate}
    \item APs, each of which will belong to an MNO, cannot use ONUs in different clouds.
    \item UEs, each of which will belong to an MNO, cannot receive services from APs connected to different clouds.
\end{enumerate}

All scenarios involve resolving three common decision problems. The primary decision in each scenario is to determine which UE will be served by which AP (\emph{AP-UE assignment}). Additionally, it must be decided which APs in the network will be active (\emph{AP-role assignment}) and how many DUs and LCs will be activated in the cloud (\emph{number of active DUs/LCs}).

In the first scenario, the service provider to which each UE belongs is known. Accordingly, the model incorporates the sets $\mathcal{A}_m$ and $\mathcal{U}_m$ for each MN $m$, where $m = 1, \ldots, M$. The decisions to be made in this scenario include the AP-UE assignment, AP-role assignment, and the number of active DUs/LCs. In line with the non-cooperation condition, each UE must be assigned only to APs that belong to its respective MN.

In the second scenario, the MNO affiliation of each AP is known, but it is initially uncertain which MNO each UE will receive service from. Accordingly, the model includes the sets $\mathcal{A}_m$, while the sets $\mathcal{U}_m$ are not defined at the outset. The decisions in this scenario involve determining the AP-UE assignment, AP-role assignment, number of active DUs/LCs, and the UE-MNO assignment. In line with the non-cooperation condition, each UE must be assigned only to APs that belong to a single MNO’s cloud and fronthaul. As a result, the UE-MNO assignment is implicitly determined: the MNO owning the APs assigned to a UE becomes the service provider for that UE.

Scenario 1 is representative of a strategy where the association between a UE and an MNO is based on a long-term contract, and a UE is not incentivized to obtain concurrent contracts with multiple MNOs (i.e., the most common practice in existing networks). 

Scenario 2 is representative of a more flexible operation strategy. It can represent a use case where a UE maintains long-term contracts with a plurality of MNOs. Alternatively it can represent a much more flexible strategy where UEs establish contracts with any of the available MNOs on an extremely short-term basis (e.g., minutes) to maximize (or minimize) a given objective function (e.g., power consumption). Indeed, blockchains are envisioned as efficient enabling technologies for such short-term dynamic contracts.

\subsection{0-1 integer programming frameworks of scenarios}

Our goal is to develop efficient deployment and provisioning schemes based on the scenarios detailed in Section~\ref{subsec:scenarios}, encompassing AP-role assignment, UE–AP associations, and MNO-specific cloud/fronthaul activation. In this subsection, we introduce our novel 0–1 integer programming problems. The solutions to these problems yield the optimal AP-UE assignment, AP-role assignment, and the number of active DUs and LCs for both Scenario 1 and Scenario 2. Additionally, in Scenario 2, the solution includes the optimal UE-MNO assignment. Table \ref{tab:Table-notation} summarizes the notation commonly used in the models for both scenarios. Note that for each MNO, the V-CRAN architecture and its parameters follow \cite{demir2022cell}. In the table, GOPS stands for giga operations per second.

\begin{table}[ht]
\centering
\vspace{-3mm}
\caption{Common symbol list }
\label{tab:Table-notation}
\resizebox{\linewidth}{!}{
\begin{tabular}{ll}
\toprule
{Sets}   &              \\ \midrule
$\mathcal{A}_{\rm all}$     & $\{1,...,L\}$ set of APs\\ 
$\mathcal{U}_{\rm all}$     & $\{1,...,K\}$ set of UEs\\
$\mathcal{N}_{\rm all}$     & $\{1,...,M\}$ set of MNOs\\
$\mathcal{DL}$     & $\{1,...,W\}$ set of DUs/LCs \\
$\mathcal{DL}_{m}$ & set of DUs / LCs belonging to MNO $m\in \mathcal{N}_{\rm all}$ \\ 
$\mathcal{A}_{m}$ & set of APs belonging to MNO $m\in \mathcal{N}_{\rm all}$ \\ \midrule
{Parameters}         &              \\ \midrule
$L$ & number of APs  \\
$K$ & number of UEs  \\
$W$ & number of DUs / LCs \\
$W_{\rm max}$ & maximum number of APs that can be assigned to one optical wavelength \\
$C_{\rm max}$ & maximum processing capacity of each DU \\
$\mathcal{X}$ & average GOPS per UE served by each active AP \\
$\mathcal{Z}$ & average GOPS per active AP \\
$\mathcal{F}$ & fixed GOPS for all UEs of desired SE \\
$\overline{C}^k_{ilr}$ & SINR constraint linearization coefficients matrix $\forall k,i\in \mathcal{U}_{\rm all},l,r\in \mathcal{A}_{\rm all}$\\
\midrule
{Decision Variables} &              \\ \midrule
$x_{kl}$ & 1 if UE $k\in \mathcal{U}_{\rm all}$ is served by AP $l\in \mathcal{A}_{\rm all}$, 0 otherwise \\
$z_{l}$ & 1 if AP $l\in \mathcal{A}_{\rm all}$ is active, 0 otherwise. \\
$l_{w}$ & 1 if LC $w\in \mathcal{DL}$ is active, 0 otherwise \\
$d_{w}$ & 1 if DU $w\in \mathcal{DL}$ is active, 0 otherwise \\
$a_{ilr}$ & 1 if UE $i\in \mathcal{U}_{\rm all}$ is served by AP $l\in \mathcal{A}_{\rm all}$ and AP $r\in \mathcal{A}_{\rm all}$, 0 otherwise \\
$\mathrm{TPC}_{m}$ & TPC of MNO $m \in \mathcal{N}_{\rm all}$, ~~$\mathrm{TPC}_{m} \geq 0$ \\
$\mathrm{TPC}^{\rm max}$ & The maximum TPC among MNOs, ~~$\mathrm{TPC}^{\rm max} \geq 0$ \\
\bottomrule
\end{tabular}}
    \vspace{-4mm}
\end{table}
To promote fairness across MNOs and avoid overloading a single operator, we minimize the maximum TPC across all MNOs rather than the total network power. Table~\ref{tab:Table-obj-coef} lists all the parameters used to compute the TPC for each MNO, along with the corresponding coefficients in the objective function.

Each MNO's cloud infrastructure is managed by a dedicated dispatcher; accordingly, the coefficient $\mathcal{C}_1$ is the power consumption of a dispatcher. An active AP consumes static power when idle, denoted by $P_{\rm AP,0}$. Since each active AP is connected to an ONU, the number of active ONUs equals the number of active APs, each contributing $P_{\rm ONU}$ to the TPC. Additionally, each active AP incurs a processing power cost in the DU, calculated as $\Delta^{\rm proc}_{\rm DU-C}\mathcal{Z}/(\sigma_{\rm cool} C_{\rm max})$; this component is captured by the coefficient $\mathcal{C}_2$.

The coefficient $\mathcal{C}_3$ denotes the power consumed by an active AP during downlink transmission to UEs. The coefficient $\mathcal{C}_4$ accounts for the power consumption of each OLT module per LC connected to a DU. Each active DU also consumes idle static power, represented by the coefficient $\mathcal{C}_5$. Since each AP can serve multiple UEs, the additional TPC in DUs resulting from a single AP-UE assignment is captured by coefficient $\mathcal{C}_6$. Finally, coefficient $\mathcal{C}_7$ represents the term related to the fixed GOPS executed by DUs to support all UEs connected to an MNO, based on the desired SE level.

\begin{table}[!htp]
\centering
\caption{Objective function coefficients}
\label{tab:Table-obj-coef}
\resizebox{\linewidth}{!}{
\renewcommand{\arraystretch}{1} 
\begin{tabular}{|c|c|c|c|}
\hline
$\mathcal{C}_1$ & $P_{\rm disp}/\sigma_{\rm cool}$ & 
$\mathcal{C}_5$ & $P^{\rm proc}_{\rm DU-C,0}/\sigma_{\rm cool}$ \\ \hline
$\mathcal{C}_2$ & $P_{\rm AP,0}+P_{\rm ONU}+\Delta^{\rm proc}_{\rm DU-C}\mathcal{Z}/(\sigma_{\rm cool}C_{\rm max})$ & 
$\mathcal{C}_6$ & $\Delta^{\rm proc}_{\rm DU-C}\mathcal{X}/(\sigma_{\rm cool}C_{\rm max})$ \\ \hline
$\mathcal{C}_3$ & $\Delta^{\rm tr}$ $p_{\rm max}$ & 
$\mathcal{C}_7$ & $\Delta^{\rm proc}_{\rm DU-C}\mathcal{F}/(\sigma_{\rm cool}C_{\rm max})$ \\ \hline
$\mathcal{C}_4$ & $P_{\rm OLT}/\sigma_{\rm cool}$ &  & \\ \hline
\end{tabular}}
\end{table}

According to each scenario, the definitions of the required sets and variables differ. Accordingly, the sets and variables in Table \ref{tab:Table-scenarios-additions} are added to the models according to the scenarios.

\begin{table}[ht]
\centering
\caption{Additions for scenarios}
\label{tab:Table-scenarios-additions}
\resizebox{\linewidth}{!}{
\begin{tabular}{lll}
\toprule
{Symbol}   & {Symbol explanation}  &   {Scenario}   \\ \midrule
$\mathcal{U}_{m}$ & set of UEs belonging to MNO $m\in \mathcal{N}_{\rm all}$ & 1 \\
$\rho_{km}$ & 1 if UE $k\in \mathcal{U}_{\rm all}$ is assigned to MNO $m\in \mathcal{N}_{\rm all}$, 0 o.w. & 2\\
\bottomrule
\end{tabular}}
\end{table}

In the following, we present the mathematical models of Scenarios 1 and 2 with their explanations, respectively.

\subsection{Scenario 1}

The mathematical model for Scenario 1, where the MNOs to which the UEs belong are known, is given below:

\allowdisplaybreaks 

\begin{align} 
&\textrm{minimize} \hspace{2mm} {\mathrm{TPC}^{\rm max}} \hspace{2mm} \label{eqn:objfnc} \\
& \textrm{s.t.} \hspace{3mm} 
\mathrm{TPC}^{\rm max} \geq \mathrm{TPC}_{m}, \hspace{5mm} \forall m \in \mathcal{N}_{\rm all} \label{eqn:TPCmax} \\
\begin{split}    
 &\mathrm{TPC}_{m}=\mathcal{C}_1 + \sum_{l \in \mathcal{A}_{m}}z_{l}(\mathcal{C}_2+\mathcal{C}_3) \\ &\hspace{13mm} + \mathcal{C}_4 \sum_{w \in \mathcal{DL}_{m}}l_{w} + \mathcal{C}_5 \sum_{w \in \mathcal{DL}_{m}}d_{w} \\ &\hspace{13mm} + \mathcal{C}_6 \sum_{l\in \mathcal{A}_{m}} \sum_{k\in \mathcal{U}_{\rm all}}x_{kl} + \mathcal{C}_7, \hspace{5mm} \forall m\in \mathcal{N}_{\rm all} \label{eqn:scenario-1-2_TPCs}
\end{split}  \\
&\sum_{l\in \mathcal{A}_{\rm all}} x_{kl} \geq 1, \hspace{5mm} \forall k\in \mathcal{U}_{\rm all} \label{eqn:UE-assignment} \\
&\sum_{k\in \mathcal{U}_{\rm all}} x_{kl} \leq K z_{l}, \hspace{5mm} \forall l\in \mathcal{A}_{\rm all} \label{eqn:scenario-1-2_AP-activate-1}\\
&z_{l} \leq \sum_{k\in \mathcal{U}_{\rm all}} x_{kl}, \hspace{5mm} \forall l\in \mathcal{A}_{\rm all} \label{eqn:scenario-1-2_AP-activate-2}\\
& a_{ilr} \leq x_{il}, \hspace{5mm} \forall i\in \mathcal{U}_{\rm all},~ l,r \in \mathcal{A}_{\rm all} \label{eqn:scenario-1-2-3_a_linear-1}\\
& a_{ilr} \leq x_{ir}, \hspace{5mm} \forall i\in \mathcal{U}_{\rm all},~ l,r \in \mathcal{A}_{\rm all} \label{eqn:scenario-1-2-3_a_linear-2}\\
& a_{ilr} \geq x_{il}+x_{ir}-1, \hspace{5mm} \forall i\in \mathcal{U}_{\rm all},~ l,r \in \mathcal{A}_{\rm all} \label{eqn:scenario-1-2-3_a_linear-3}\\
&\sum_{l\in \mathcal{A}_{m}} z_{l} \leq W_{\rm max}|\mathcal{DL}_{m}|, \hspace{5mm} \forall m\in \mathcal{N}_{\rm all} \label{eqn:scenario-1-2_APnum-limit}\\
&\sum_{w\in \mathcal{DL}_{m}} l_{w} \geq \sum_{l\in \mathcal{A}_{m}} z_{l}/W_{\rm max}, \hspace{5mm} \forall m\in \mathcal{N}_{\rm all} \label{eqn:scenario-1-2_LCnum}\\
\begin{split} 
&\mathcal{Z} \sum_{l\in \mathcal{A}_{m}} z_{l} + \mathcal{X} \sum_{k\in \mathcal{U}_{\rm all}}\sum_{l\in \mathcal{A}_{m}} x_{kl} \\ &\hspace{20mm} + \mathcal{F} \leq C_{\rm max}\sum_{w\in \mathcal{DL}_{m}} d_{w}, \hspace{5mm} \forall m\in \mathcal{N}_{\rm all} \label{eqn:scenario-1-2_GOPS-limit}
\end{split}  \\
&\sum_{w\in \mathcal{DL}_{m}} l_{w} \leq \sum_{w\in \mathcal{DL}_{m}} d_{w}, \hspace{5mm}  \forall m\in \mathcal{N}_{\rm all} \label{eqn:LCnum-DUnum-limit}\\
&\sum_{i\in \mathcal{U}_{m}}\sum_{l\in \mathcal{A}_{m}}\sum_{r\in \mathcal{A}_{m}} a_{ilr}\overline{C}^{k}_{ilr} + 1 \leq 0, ~ \forall m\in \mathcal{N}_{\rm all}, k \in \mathcal{U}_{m} \label{eqn:scenario-1_SINR-cons-linearization}\\
& x_{kl} \leq 0, \hspace{5mm} \forall k\in \mathcal{U}_{m},~ l \in \mathcal{A}_{t},~ m,t \in \mathcal{N}_{\rm all} : m \neq t \label{eqn:scenario-1_noncooperation-cond}\\
\begin{split} 
& x_{kl}, z_{l}, l_{w}, d_{w}, a_{klr} \in \left\{0,1\right\}, \\ &\hspace{25mm} \forall k\in \mathcal{U}_{\rm all},~ w \in \mathcal{DL},~l,r\in \mathcal{A}_{\rm all}. \label{eqn:scenario-1_sign-cons}
\end{split}
\end{align}

The objective function of our 0-1 integer programming model is to minimize the power of the MNO that consumes the most power in total \eqref{eqn:objfnc}. The TPC of the MNO that consumes the maximum power is found with the constraint \eqref{eqn:TPCmax}. \eqref{eqn:scenario-1-2_TPCs} calculates the TPC of each MNO. \eqref{eqn:UE-assignment} guarantees that each UE is assigned to at least one AP. \eqref{eqn:scenario-1-2_AP-activate-1} ensures that no UE is assigned to an inactive AP, while \eqref{eqn:scenario-1-2_AP-activate-2} prevents an AP with no UE assigned to it from becoming active. The constraints \eqref{eqn:scenario-1-2-3_a_linear-1}–\eqref{eqn:scenario-1-2-3_a_linear-3} are added to the model to include the relations $a_{ilr}=x_{il}x_{ir}$ using linear inequalities. \eqref{eqn:scenario-1-2_APnum-limit} limits the total number of active APs belonging to the MNO according to the maximum number of APs that can be assigned to all DUs belonging to the same MNO. \eqref{eqn:scenario-1-2_LCnum} ensures that the minimum required number of LCs is active, calculated according to the number of active APs for each MNO. \eqref{eqn:scenario-1-2_GOPS-limit} guarantees that the total GOPS in the cloud of each MNO does not exceed the processing capacity of all active DUs in it. The constraint \eqref{eqn:LCnum-DUnum-limit} ensures that the total number of active LCs does not exceed the total number of active DUs for each MNO. \eqref{eqn:scenario-1_SINR-cons-linearization} guarantees that the minimum SINR requirement resulting from the assignment of each UE to APs is met within its MNO. According to the rule that MNOs do not cooperate, each UE is assigned to an AP belonging to its MNO with the constraint \eqref{eqn:scenario-1_noncooperation-cond}. The model demonstrates the binary variables via \eqref{eqn:scenario-1_sign-cons}.

\subsection{Scenario 2}

The mathematical model for Scenario 2, where the MNOs to which the APs belong are known but the UEs are unknown, is given below:
\allowdisplaybreaks  
\begin{align}
&\textrm{Objective function:} \hspace{2mm} \eqref{eqn:objfnc} &&\nonumber\\
& \textrm{s.t.} \hspace{3mm} 
\eqref{eqn:TPCmax}, \eqref{eqn:scenario-1-2_TPCs}, \eqref{eqn:UE-assignment}, \eqref{eqn:scenario-1-2_AP-activate-1}, \eqref{eqn:scenario-1-2_AP-activate-2}, \eqref{eqn:scenario-1-2-3_a_linear-1}–\eqref{eqn:scenario-1-2-3_a_linear-3}, \eqref{eqn:scenario-1-2_APnum-limit}, \eqref{eqn:scenario-1-2_LCnum}, \eqref{eqn:scenario-1-2_GOPS-limit}, \eqref{eqn:LCnum-DUnum-limit} &&\nonumber\\
\begin{split}
&\sum_{i\in \mathcal{U}_{\rm all}}\sum_{l\in \mathcal{A}_{m}}\sum_{r\in \mathcal{A}_{m}} a_{ilr}\overline{C}^{k}_{ilr} + 1 \leq \mathcal{M} (1-\rho_{km}), \\ &\hspace{45mm} \forall m\in \mathcal{N}_{\rm all},~ k \in \mathcal{U}_{\rm all} \label{eqn:scenario-2_SINR-cons-linearization}
\end{split} \\
\begin{split}
& x_{kl} + x_{kr} \leq 1, \\ &\hspace{5mm} \forall k\in \mathcal{U}_{\rm all},~ l \in \mathcal{A}_{m},~ r \in \mathcal{A}_{t},~ m,t \in \mathcal{N}_{\rm all} : m \neq t \label{eqn:scenario-2_noncooperation-cond}
\end{split} \\
& \rho_{km} \geq x_{kl}, \hspace{10mm} \forall k\in \mathcal{U}_{\rm all},~ l \in \mathcal{A}_{m},~ m \in \mathcal{N}_{\rm all} \label{eqn:scenario-2_rho-calc-1} \\
& \rho_{km} \leq \sum_{l\in \mathcal{A}_{m}} x_{kl}, \hspace{10mm} \forall k\in \mathcal{U}_{\rm all},~ m \in \mathcal{N}_{\rm all} \label{eqn:scenario-2_rho-calc-2} \\
\begin{split} 
& x_{kl}, z_{l}, l_{w}, d_{w}, a_{klr}, \rho_{km} \in \left\{0,1\right\}, \\ &\hspace{10mm} \forall k\in \mathcal{U}_{\rm all},~ w \in \mathcal{DL},~l,r\in \mathcal{A}_{\rm all},m\in \mathcal{N}_{\rm all}. \label{eqn:scenario-2_sign-cons}
\end{split}
\end{align}

In the second model, unlike the first scenario, some updates are made due to the absence of the $\mathcal{U}_{m}$ set.
The constraint \eqref{eqn:scenario-1_SINR-cons-linearization}, which guarantees that the minimum SINR requirement for each UE is met, is updated with \eqref{eqn:scenario-2_SINR-cons-linearization}, where $\mathcal{M}$ is the big-M parameter—a sufficiently large constant used in integer programming to activate or deactivate constraints. The constraint \eqref{eqn:scenario-1_noncooperation-cond} that ensures that each UE is assigned to an AP belonging to its MNO is replaced with \eqref{eqn:scenario-2_noncooperation-cond}. \eqref{eqn:scenario-2_rho-calc-1} and \eqref{eqn:scenario-2_rho-calc-2} ensures that the variable $\rho_{km}$, which indicates the MNO to which UEs are assigned, takes correct values as a result of the assignment of UE-AP. Finally, the binary variables that demonstrate the model are shown with \eqref{eqn:scenario-2_sign-cons}.

\section{Numerical Results and Discussion} \label{sec:analysis}
In our study, the parameter space created to obtain the results of the mathematical models we established based on various scenarios is given in Section~\ref{subsec:exp-setup}. The results obtained are detailed in Section~\ref{subsec:scenario-results}.

\subsection{Experimental setup} \label{subsec:exp-setup}
In this section, we present the system parameters that are listed in Table~\ref{tab:Table-notation} using those in Table~\ref{tab:simulation}. Each MNO in the system has its own cloud, and accordingly the number of DUs\&LCs in each cloud is $W/M$. Likewise, the number of APs in the network is equally shared among the MNOs as $L/M$. For the first scenario, the UEs are shared between MNOs equally. The AP-UE channels are generated following the methodology in \cite{demir2022cell}, and we assume that all UEs belong to the same MNO when computing the LP-MMSE precoding vectors. Although this represents a worst-case assumption, it is necessary for the implementation in Scenario 2, where the UE-MNO associations are initially unknown.

We consider a $4 \times 4$ grid-shaped deployment area and generate 35 random instances with $M = 2$ MNOs. In each instance, $K$ UEs are randomly placed within the area, while APs are positioned at all center points of the grid, which serve as potential locations for active APs. These potential AP positions are partitioned among the APs based on a homogeneous spatial distribution of the MNOs. The fractional power allocation is used as described in \cite[Eqn. (6.36)]{cell-free-book}.

We calculate the big-M value $\mathcal{M}$ in Scenario 2 as the sum of positive terms in $\overline{C}^k_{ilr}$. In this way, we obtain an upper bound that will not affect the relevant constraints in a particular case. As a result, the $\mathcal{M}$ value is calculated as 140. The GOPS parameters $\mathcal{X}$, $\mathcal{Z}$, and $\mathcal{F}$ are computed according to \cite{demir2022cell}.

All mathematical problems are solved optimally using Python 3.8 with Gurobi 11.0.2 solver on a computer with 64 GB of RAM, 12th Gen Intel(R) Core(TM) i7-12700K @ 3.60 GHz, and 475 GB of disk space. MATLAB and Python are used to visualize the test results.

\subsection{Results of scenarios 1 and 2} \label{subsec:scenario-results}

\begin{figure}[!ht]
    \centering
    \includegraphics[width=0.4\textwidth, trim=15 0 30 20, clip]{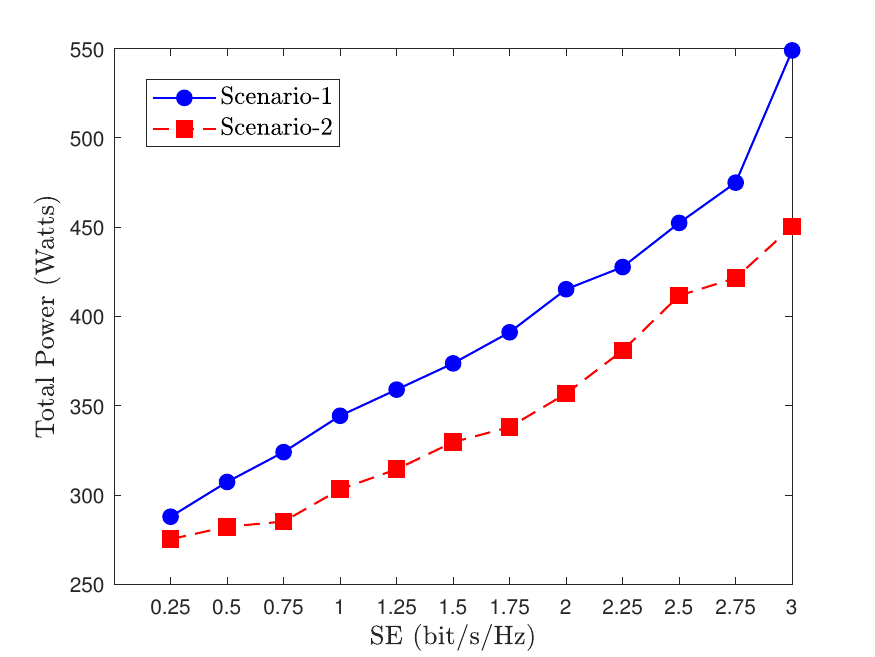}
        \vspace{-2mm}
    \caption{The average TPC vs. SE.}
\label{fig:scenarios-TPCvsSE}
    \vspace{-2mm}
\end{figure}

We generate 35 random common instances for each  SE value, and each data point in Fig.~\ref{fig:scenarios-TPCvsSE} represents the average TPC across these instances. The average time to obtain the solution per data point, considering all SE values in the range $[0.25, 3]$, is less than 1 second for Scenario 1 and between 0 and 6 seconds for Scenario 2. We emphasize that all presented results correspond to optimal solutions for the respective instances.

\begin{figure*}[t]   
	\centering
	\begin{tabular}{cc}
	\subfloat[PC vs. SE for Scenario 1]{
	{\includegraphics[width=0.4\textwidth, trim=10 0 30 20, clip]{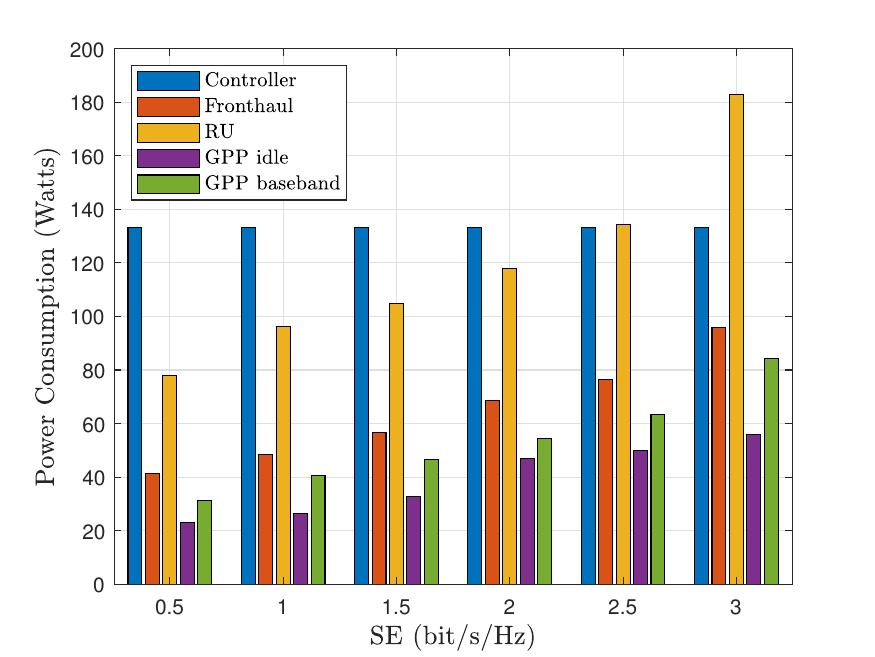}\label{fig:PCvsSE_sc1:a}}} &
	\subfloat[PC vs. SE for Scenario 2]{
	{\includegraphics[width=0.4\textwidth, trim=10 0 30 20, clip]{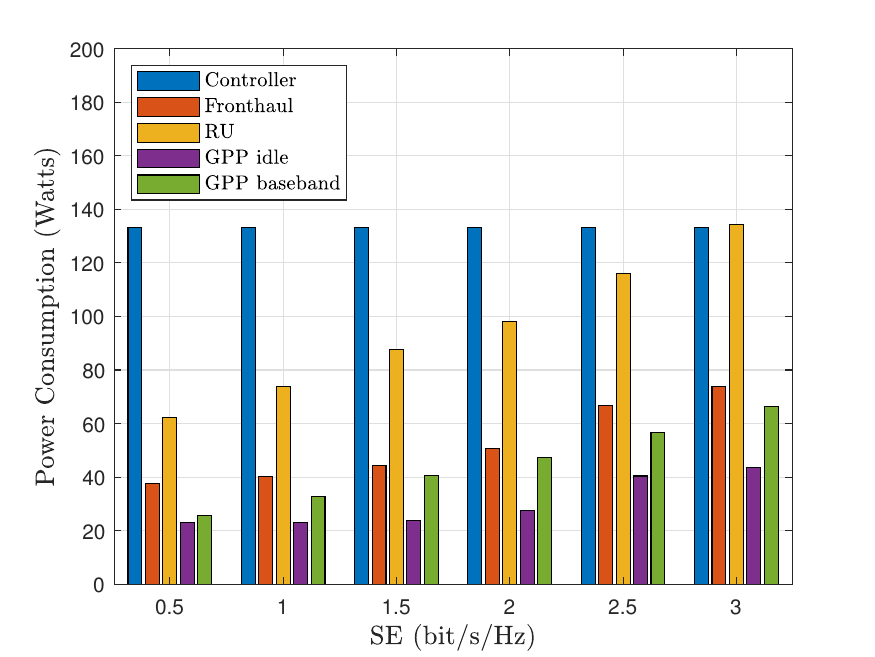}\label{fig:PCvsSE_sc2:b}}} 
	\end{tabular}
        \vspace{-2mm}
	\caption[] {The average power consumption breakdown for different SE values of the scenarios.}
	\label{fig:scenarios-PCvsSE}
    \vspace{-4mm}
\end{figure*}

In Fig.~\ref{fig:scenarios-TPCvsSE}, the solutions are compared using the same set of instances. The results for all scenarios show that the TPC in the network increases with higher SE values. Unlike Scenario 1, Scenario 2 does not include the set $\mathcal{U}_m$, meaning that the UE–MNO assignment is determined as part of the optimization. As a more flexible model, Scenario 2 consistently yields better performance (less TPC) across all SE values. The significant reduction in power consumption highlights the energy-saving potential of allowing flexible UE-MNO assignment.

To gain deeper insight into the system behavior, we analyze the power consumption breakdown in Fig.~\ref{fig:scenarios-PCvsSE} for both scenarios. The blue bar represents the fixed power consumed by the cloud dispatcher, which remains constant. As the SE requirement increases, more APs are activated, along with additional DUs and LCs. This results in higher power consumption in the radio units (RUs), fronthaul, GPPs. However, this increase is less pronounced under flexible UE–MNO assignment, as it leads to a smaller number of activated network components. In particular, when the SE requirement is 3\,bit/s/Hz for all UEs in the network, the RU power consumption in Scenario 2 is significantly lower compared to Scenario 1.

\begin{table}[!ht]
	\footnotesize
		\vspace{-0.2cm}
	\caption{Simulation Parameters.}  \label{tab:simulation}
	\centering
	\begin{tabular}{|c|c|c|c|}
		\hline
	 $L$, $K$, $W$, $N$&   16, 8, 8, 4    &    
	 $f_s$ &  30.72\,MHz    \\ \hline
	 $B$  & 20\,MHz    &
	 $N_{\rm DFT}$, $N_{\rm used}$ & 2048, 1200 \\   \hline
	 $T_s$ & 71.4\,$\mu$s &
	 $N_{\rm smooth}$, $N_{\rm slot}$ & 12, 16 \\ \hline
	 $\tau_c$, $\tau_p$ & 192, 8 &
Size of coverage area &  1\,km $\times$\,1\,km      \\ 
	 \hline 
	 $P_{{\rm AP},0}$, \ $\Delta^{\rm tr}$  &  6.8$N$\,W, \ 4  &    
	  Pilot power, \ $p_{\rm max}$ & 100\,mW, \   1\,W    \\ 
	 \hline  
 $P_{\rm disp}$, \ $\sigma_{\rm cool}$  & 120\,W, \ 0.9 &
  $P_{\rm ONU}$, \ $P_{\rm OLT}$ & 7.7\,W, \ 20\,W  \\  \hline
 $P_{\rm DU-C,0}^{\rm proc}$ & 20.8\,W &
 $\Delta^{\rm proc}_{\rm DU-C}$ & 74\,W \\ \hline
 $C_{ \rm max}$ & 180\,GOPS &
 $R_{\rm max}$, $N_{\rm bits}$ & 10\,Gbps, 12 \\ \hline
	\end{tabular}
\vspace{-5mm}
\end{table}

\section{Conclusions}

This paper presents a holistic optimization framework for energy-aware service provisioning in multi-operator cell-free massive MIMO networks deployed over V-CRAN architectures. Two distinct deployment scenarios were investigated, differing in their constraints on UE–MNO assignment flexibility. Through novel 0–1 integer programming formulations, we jointly optimized AP selection, UE association, and cloud resource allocation to minimize the maximum TPC among operators.

Simulation results reveal that enabling flexible UE–MNO assignment, as in Scenario 2, yields substantial energy savings compared to the static assignment model of Scenario 1. These savings become increasingly pronounced at higher SE requirements, where resource demands intensify. The power breakdown analysis further demonstrates that flexibility in UE–MNO association reduces the number of active network components—particularly RUs—thus mitigating the growth in TPC.

Overall, the results highlight the significant potential of dynamic UE–MNO assignment in improving energy efficiency in cell-free V-CRAN systems. These findings provide valuable insights for designing sustainable and resource-efficient architectures in future 6G networks.

\bibliographystyle{IEEEtran}

\bibliography{IEEEabrv,refs}

\end{document}